\documentstyle[twocolumn,aps,epsf]{revtex}

\newcommand{\beq}{\begin{equation}}
\newcommand{\eeq}{\end{equation}}
\newcommand{\beqar}{\begin{eqnarray}}
\newcommand{\eeqar}{\end{eqnarray}}
\input{psfig.sty}

\begin{document}
\draft
\wideabs
{
\title{Detection of Topological Patterns in Complex Networks: \\
Correlation Profile of the Internet}
\author{Sergei Maslov$^1$, Kim Sneppen$^{2,3}$, Alexei Zaliznyak$^1$}
\address{$^1$ Department of Physics, Brookhaven National Laboratory,
Upton, New York 11973, USA.}
\address{$^2$ Niels Bohr Institute, Blegdamsvej 17,
DK - 2100, Copenhagen, Denmark.}
\address{$^3$ Department of Physics,
Norwegian University of Science and Technology,
N-7491 Trondheim, Norway.}

\date{\today}
\maketitle
\begin{abstract}
A general scheme for detecting and analyzing topological
patterns in large complex networks is presented.
In this scheme the network in question
is compared with its properly randomized version
that preserves some of its low-level
topological properties.
Statistically significant deviation
of any measurable property of a network
from this null model likely
reflect its design principles
and/or evolutionary history.
We illustrate this basic scheme on the example of
the correlation profile of the Internet
quantifying correlations between connectivities of
its neighboring nodes. This profile
distinguishes the Internet from previously studied
molecular networks with a similar scale-free connectivity
distribution. We finally demonstrate that
clustering in a network is
very sensitive to both the connectivity distribution and
its correlation profile and compare the clustering
in the Internet to the appropriate null model.
\end{abstract}
\pacs{89.75.-k, 89.70.+c, 05.10.-a, 05.65.+b}
%\twocolumn]
Keywords: Random networks, scale free networks,
correlation profile, cliquishness,
metropolis, network motifs.
}
\narrowtext

Networks have emerged as a unifying theme in
complex systems research. It is in fact no
coincidence that networks and complexity are
so heavily intertwined. Any future definition
of a complex system should reflect the fact that
such systems consist of many mutually interacting
components. These components are not
identical as say electrons in condensed matter physics.
Instead each of them has a unique identity separating
it from others. The very basic
question one may ask about a complex system is which other components
a given component interacts with? Systemwide this information
can be visualized as a graph whose nodes correspond to individual
components and edges to their mutual interactions.
Such a network can be thought of as a backbone of the
complex system along which propagate various
signals and perturbations.

Living organisms provide us with a quintessential paradigm for a
complex system. Therefore, it should not be surprising that
in biology networks appear on many different levels:
from genetic regulation and signal transduction in
individual cells, to neural system of animals, and finally to
food webs in ecosystems. However, complex networks are not
limited to living systems: in fact
they lie at the foundation of an increasing number of artificial
systems. The most prominent example of this is the Internet and the
World Wide Web being correspondingly
the ``hardware'' and the ``software'' of the
network of communications between computers.

An interesting common feature of many complex networks
is an extremely broad, often scale-free, distribution of connectivities
(defined as the number of immediate neighbors) of their nodes
\cite{barabasi_scale_free}. While the majority of nodes in such
networks are each connected to just a handful of neighbors,
there exist a few hub nodes that have a disproportionately large
number of interaction partners. The histogram of connectivities
is an example of a low-level topological property of a network.
While it answers the question
about how many neighbors a given node has, it gives
no information about the identity of those neighbors. It is clear that
most of non-trivial properties of networks lie in the exact
way their nodes are connected to each other.
However, such connectivity patterns are rather difficult
to quantify and measure.
By just looking at many large complex networks one gets the impression
that they are wired in a rather haphazard way.
One may wonder which topological properties of a given network
are indeed random, and which arose due to evolution and/or fundamental
design principles and limitations? Such non-random features
can then be used to identify the network
and better understand the underlying
complex system.

In this work we propose a universal recipe for how such
information can be extracted.
To this end we first construct a proper null randomized model of a given
network. As was pointed out in \cite{newman},
broad distributions of connectivities in most
real complex networks indicate that the connectivity
is an important individual characteristic of a node and as such
it should be preserved in any meaningful randomization
process. In addition to connectivities one may choose to
preserve some other low-level topological properties
of the network. Any higher level topological property,
such as e.g. the pattern of correlations between connectivities
of neighboring nodes,
the number of loops of a certain type, the number and sizes
of components, the diameter of the network, spectral properties of
its adjacency matrix, can then be measured in the real complex network
and separately in an ensemble of its randomized counterparts.
Dealing with the whole ensemble
allows one to put error bars on any quantity measured
in the randomized network. One then concentrates only on those topological
properties of the complex network that significantly deviate from
the null model, and, therefore, are likely
to reflect its basic design principles
and/or evolutionary history.

\begin{figure}[t]
\centerline{\psfig{file=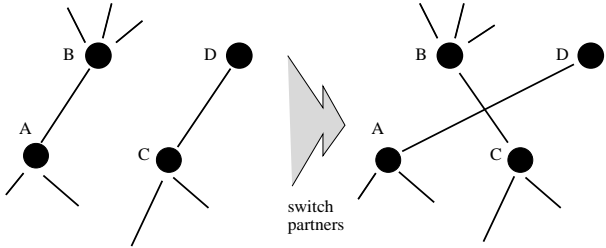,angle=0,width=8cm}}
\vspace{0.5cm}
\caption[]
{One elementary step of the local rewiring
algorithm. A pair of
edges A---B and C---D
is randomly selected. They are then rewired
in such a way that A becomes connected to D, and C  - to B,
provided that none of these edges already exist in the network, in
which case the rewiring step is aborted, and a new pair of edges is
selected.
The last restriction prevents the appearance of multiple
edges connecting the same pair of nodes.
}
\label{switch}
\end{figure}

The {\it local rewiring algorithm} that randomizes a network yet
strictly conserves connectivities of its nodes
\cite{maslov_sneppen_science,math_rewire}
consists of repeated application of the
elementary rewiring step shown and explained in detail
in Fig.\ref{switch}.
It is easy to see that the number of neighbors of every node in
the network remains unchanged after an elementary step of this
randomization procedure. The directed network version of this
algorithm separately conserves the number of upstream and
downstream neighbors (in- and out-degrees) of every node.

Another simple numerical algorithm
generating  such a random network ``from scratch''
was proposed in \cite{newman,bender}.
It starts with assigning to each node a number
$k_i$ of ``edge stubs'' equal to its desired connectivity.
A random network is then constructed by randomly picking two
such edge stubs and joining them together to form a real edge
connecting these two nodes.
One of the limitations of this ``stub reconnection'' algorithm is that
for broad distribution of connectivities, which is
usually the case in complex networks \cite{barabasi_scale_free},
the algorithm generates multiple edges joining the same pair
of hub nodes. This problem cannot be avoided by simply
not allowing multiple edges
to form during the reconnection process
as in this case the whole algorithm
would get stuck in a configuration in which
the remaining edge stubs have no
eligible partners. Fortunately the local rewiring
algorithm \cite{maslov_sneppen_science,math_rewire}
instead of completely deconstructing a network and then
randomly putting it back together, only gradually changes
its wiring pattern. Hence, any topological
constraint such as e.g. that of no multiple edges, or no
disconnected components, can be maintained
at each step of the way.

Once an ensemble of randomized versions of a
given complex network is generated, the abundance
of any topological pattern is compared between
the real network and characteristic members of this ensemble.
This comparison can be quantified using two natural parameters:
1) the ratio
$R(j)=N(j)/\overline{N_r(j)}$,
where $N(j)$ is the number of times the pattern $j$ is observed
in the real network, and  $\overline{N_r(j)}$ is the average number
of its occurrences in the ensemble of its random counterparts;
2) the Z-score of the deviation defined as
$Z(j)=[N(j)-\overline{N_r(j)}]/\Delta N_r(j)$,
where $\Delta N_r(j)$
is the standard deviation of
$N_r(j)$ in the randomized ensemble.
This general idea was recently applied to
protein networks in yeast \cite{maslov_sneppen_science}
and {\sl E. coli} \cite{uri}.

We now illustrate our general methods using
the example of the Internet, defined on the
level of Autonomous Systems (AS).
Autonomous Systems are large groups of
workstations, servers, and routers usually
belonging to one organization such as e.g.
a university, or a business enterprise.
The data on direct connections between Autonomous Systems
is regularly updated and is available on the website
of the National Laboratory for Applied Network Research
\cite{nlanr}.
Such coarse-grained structure of the Internet was
a subject of several recent studies
\cite{Faloustos,Vespignani2001,Yook2002,Goh2002}.
In the following analysis we use the millennium snapshot of the Internet
(data from January 2, 2000), when $N=6474$ Autonomous Systems
were linked by $E=12572$ bi-directional edges.

It was recently reported \cite{Faloustos}
that the Internet is characterized by a scale-free
distribution of AS connectivities
$p(K) \propto 1/K^{\gamma}= 1/K^{2.1\pm0.2}$.
One can show that for such a scale-free network
the above mentioned constraint of
no multiple connections between nodes is
extremely important. Indeed, the connectivity of two
largest connected hubs in a scale-free networks scales
as $k_{max} \sim N^{1/(\gamma-1)}$. In an uncorrelated random
network with no constraints on edge multiplicity
the expected number of edges connecting these two hubs
scales as $k_{max}^2/(2E) \sim N^{2/(\gamma-1)-1}$ and increases
indefinitely for $\gamma<3$ (here we assumed that
$E \sim N$). For the Internet that
corresponds to two largest  hubs
with connectivities of respectively $K_0=1458$
and $K_1=750$ being connected by a swooping
$K_0 K_1/ (2E)=1458 \cdot 750/(2 \cdot 12572)=43.5$ edges!
Hence, in this case a random network ensemble generated by
our local rewiring algorithm is very
different from the one generated by the stub
reconnection algorithm and analytically studied
in \cite{newman}.

\begin{figure}[t]
\centerline{\psfig{file=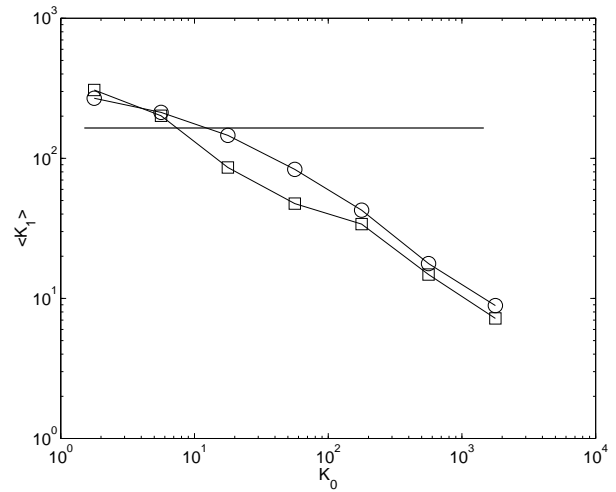,angle=0,width=8cm}}
\vspace{0.5cm}
\caption[]
{The average connectivity
$\langle K_1 \rangle_{K_0}$
of neighbors of nodes with
connectivity $K_0$ in the Internet (squares) and its typical randomized
counterpart (circles). Error bars in multiple realizations of the randomized network are
smaller than symbol sizes. The horizontal line is the analytical result
$\langle K_1 \rangle_{K_0}={\rm const}= \langle K^2 \rangle/\langle K
\rangle \simeq 165$ valid for a random network in which
multiple edges between pairs of nodes are allowed \cite{newman}.}
\label{fig_K1_vs_K0}
\end{figure}

Fig.\ref{fig_K1_vs_K0} shows
the average connectivity
$\langle K_1 \rangle_{K_0}$ of neighbors
of nodes with the connectivity $K_0$
in the real Internet (squares)
as well as in a typical random network with no
multiple connections between nodes
generated by our local rewiring algorithm
(circles). From this figure it is  clear that most
of the $\langle K_1 \rangle_{K_0} \propto K_0^{-0.5}$
dependence reported in Ref.  \cite{Vespignani2001}
is reproduced in our random ensemble and hence
can be attributed to the effective
repulsion between hubs due to
the constraint of having no more than one
edge directly connecting them to each other.
In the absence of correlations between node connectivities
by definition $\langle K_1 \rangle_{K_0}={\rm const}= \langle K^2 \rangle/\langle K
\rangle$ \cite{newman}. This expression,
shown as a horizontal line in Fig.\ref{fig_K1_vs_K0},
applies only to a randomized network in which multiple edges
are allowed. In an ensemble of random scale-free networks with no
multiple edges the conditional probability
distribution $P(K_1|K_0)$ crosses over between
$K_1/K_1^{\gamma}$ for $K_1 \ll K_1^*=2E/K_0$ to
$1/K_1^{\gamma}$ power law tail for
$K_1 \gg  K_1^*$. This makes
$\langle K_1 \rangle_{K_0}$ to
asymptotically scale as $K_0^{\gamma-3}$.
We have confirmed numerically that
$P(K_1|K_0)$ in our randomized ensemble
has a very similar shape to that observed in the
real Internet \cite{Goh2002}.

\begin{figure}[t]
\centerline{\psfig{file=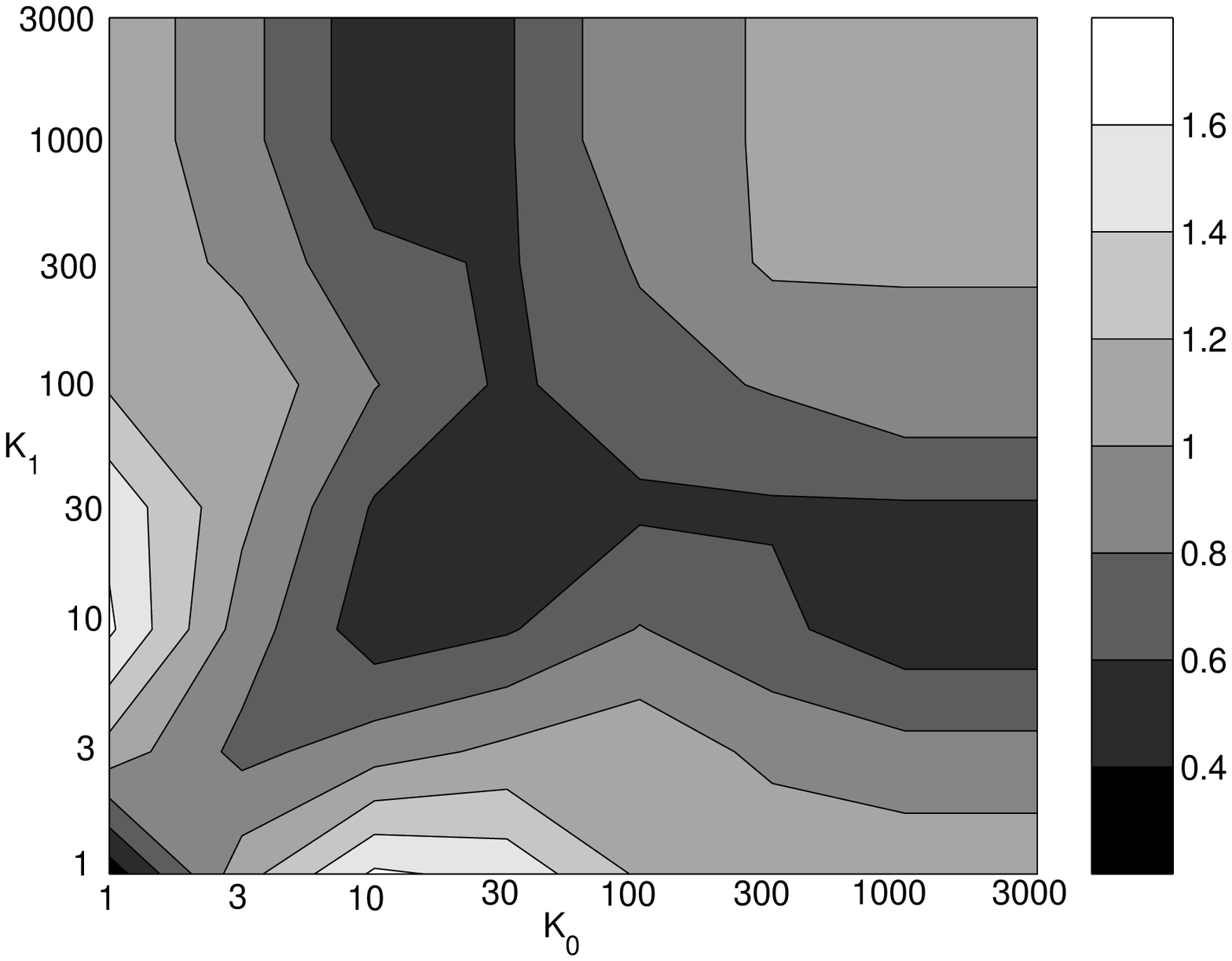,angle=0,width=8cm}}
\vspace{0.5cm}
\caption[]
{Correlation profile of the Internet.
The ratio $R(K_0,K_1)=N(K_{0},K_{1})/\overline{N_r(K_{0},K_{1})}$,
where $N(K_{0},K_{1})$ is the total number of edges in the
Internet connecting pairs of Autonomous Systems with
connectivities $K_{0}$ and $K_{1}$,
while $\overline{N_r(K_{0},K_{1})}$ is the same quantity
in the ensemble of randomized versions of the Internet, generated by the
local rewiring algorithm described in the text.}
\label{fig_R_internet}
\end{figure}

\begin{figure}[t]
\centerline{\psfig{file=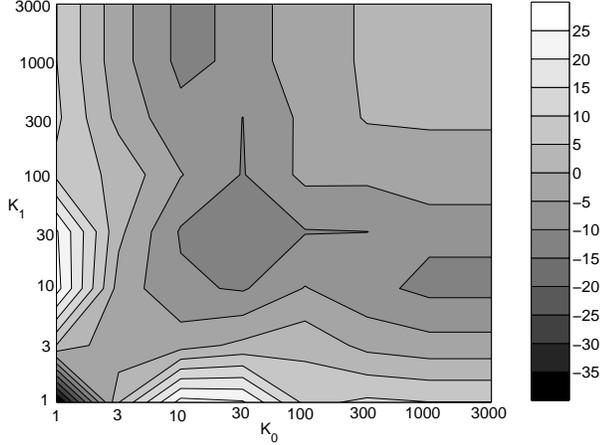,angle=0,width=8cm}}
\vspace{0.5cm}
\caption[]
{Statistical significance of correlations in the Internet.
The Z-score of correlation patterns in the internet
$Z(K_0,K_1)=(N(K_0,K_1)-\overline{N_r(K_0,K_1)})/\Delta N_r(K_0,K_1)$.
Here $\Delta N_r(K_0,K_1)$ is the standard deviation of $N_r(K_0,K_1)$
measured in an ensemble of 1000 randomized networks.}
\label{fig_Z_internet}
\end{figure}

From the above discussion one may get the impression
that the topology of the Internet is in perfect
agreement with its randomized version.
This is however not true. Let $N(K_0,K_1)$ to denote the total
number of edges connecting nodes with
connectivities $K_0$, and $K_1$. This is an example
of a higher level topological property of a complex network, which
can be compared to its typical value $\overline{N_r(K_0,K_1)}$
in the appropriate null-model network.
By comparing $N(K_0,K_1)$ and $\overline{N_r(K_0,K_1)}$ one
measures the {\it correlation profile} of the complex network,
formed by correlations in connectivities of neighboring nodes.
In Fig.\ref{fig_R_internet} we visualize the
correlation profile of the Internet by plotting the ratio
$R(K_0,K_1)=N(K_0,K_1)/\overline{N_r(K_0,K_1)}$.
Regions on the $K_0-K_1$ plane, where $R(K_0,K_1)$
is above (below) 1 correspond to enhanced (suppressed)
connections between nodes with these connectivities in
the Internet compared to its randomized counterpart.
The statistical significance of these deviations,
measured by the Z-score
$Z(K_0,K_1)=(N(K_0,K_1)-\overline{N_r(K_0,K_1)})/\Delta
N_r(K_0,K_1)$,
is shown in Fig.\ref{fig_Z_internet}
Our analysis is based on an ensemble of
1000 randomized networks with connectivities
logarithmically binned into two bins per decade.
In Figs.\ref{fig_R_internet},\ref{fig_Z_internet} one can see several
prominent features:
\begin{itemize}
\item Strong suppression of edges between
nodes of low connectivity $3 \geq K_0,K_1 \geq 1$.
\item Suppression of edges between nodes that both are of
intermediate connectivity $100>K_0,K_1 \geq 10$,
\item Strong enhancement of the number of edges connecting
nodes of low connectivity $3 \geq K_0 \geq 1$
to those with intermediate connectivity $100>K_1 \geq 10$.
\end{itemize}
On the other hand any pair among 5 hub nodes with
$K_0, K_1 > 300$ was found to be connected by an edge, both
in the real network, and in a typical random sample.
Hence $R(K_0,K_1)$ is close to 1 in the upper right
corner of Fig.\ref{fig_R_internet}.

The strong suppression of connections between pairs of
nodes of low connectivity can in part be
attributed to the constraint that all AS on the
Internet have to be connected
to each other by at least one path.
We have explicitly checked that there are indeed
no isolated clusters in our data for the Internet.
However, when we used an ensemble of
random networks in which the formation of
isolated clusters was prevented at every rewiring step,
we found very little change in the observed
correlation profile.
The division of all nodes on the Internet
into three distinct groups of low-, intermediate-,
and highly-connected ones  visible in its correlation profile
may be due to its hierarchical structure of, correspondingly,
users, low-level (possibly regional) Internet Service Providers
(ISP), and high-level (global)
ISP. Similar hierarchical picture was recently suggested in
Ref. \cite{Capocci2001} on the basis of the traceroute data.

It is worthwhile to note that the correlation profile of the
Internet measured in this work makes it qualitatively
different from yeast protein networks analyzed by us earlier
\cite{maslov_sneppen_science}.
Those molecular networks are characterized by
suppressed connections between nodes of very high connectivity,
and increased number of links between nodes of
intermediate connectivity. Thus correlation profile allows
one to differentiate between otherwise very similar
scale-free networks in various complex systems.

The correlation profile is by no means the only topological
pattern one can investigate in a given complex network, with other
examples being its spectral dimension \cite{Bilke2002},
the betweenness of its edges and nodes \cite{girvan_newman,Vespignani2001},
feedback, feed-forward loops, and other small network motifs
\cite{uri}. In the rest of
this paper we analyze the level of
clustering \cite{watts} of the Internet,
quantified by its number of loops of length 3 (triangles).
The real Internet contains $6584$ such loops, while
its random counterparts, generated by our local rewiring
algorithm, have $8636 \pm 224$ triangles (this and all future results were
measured in an ensemble of 100 randomized networks.) Thus the
clustering of the real Internet is some 9 standard deviations
{\it below} its value in a randomized network! This result is
surprising because there are
good reasons for the Internet to have above average level of clustering.
Indeed, one expects its nodes to preferentially link according to
their geographical location \cite{Vespignani2001,Yook2002},
general type of business or academic
enterprises they represent, etc. All these factors usually tend to
increase clustering \cite{watts}. On the other hand, the
correlation profile of the Internet visualized in
Fig.\ref{fig_R_internet} naturally leads to the reduction in
clustering. Indeed, the suppression of connections between nodes
of intermediate connectivity in favor of nodes of low connectivity
should reduce the number of triangles in the network.

\begin{figure}[t]
\centerline{\psfig{file=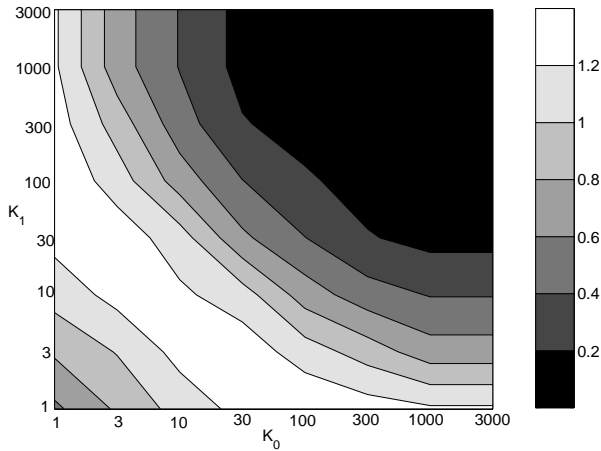,angle=0,width=8cm}}
\vspace{0.5cm}
\caption[]
{The correlation profile $R(K_0,K_1)$ of a network with the same
set of connectivities as the Internet but with no triangles.
Note the suppression of connections between different
hubs in favor of connections between hubs and nodes of low
connectivity.}
\label{fig_R_no_triangles}
\end{figure}

\begin{figure}[t]
\centerline{\psfig{file=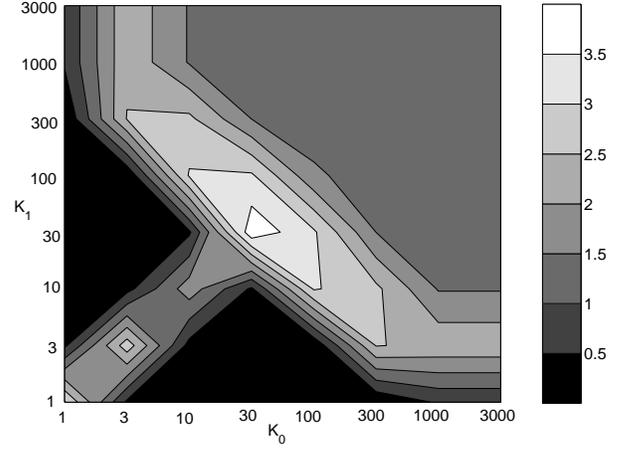,angle=0,width=8cm}}
\vspace{0.5cm}
\caption[]
{The correlation profile $R(K_0,K_1)$ of a network with the same
set of connectivities as the Internet but with a very large
number triangles (59144).
Note the tendency of nodes with similar connectivities to connect
to each other.}
\label{fig_R_60000_triangles}
\end{figure}

In order to explore the interplay between the level of clustering
in the network and its correlation profile we studied two
``extremal'' random networks with the same connectivities of nodes
as the real Internet. The first network contained no triangles,
while the second one had a swooping 59144 triangles. Both networks were
generated using a simple modification of our basic local rewiring algorithm
in which a rewiring step was accepted only if it did not increase
(in the first case) or decrease (in the second case)
the number of triangles in the network. In the first case after some transient time
all triangles have disappeared from the network, at which point we
measured its correlation profile (Fig.\ref{fig_R_no_triangles}).
In the second case our algorithm was designed to generate a network with the
largest possible number of triangles . Computer time limitations have forced us
to stop the program when we reached 59144 triangles, which as will be shown later is
rather close to the absolute maximum of 63844 triangles for a given set of node
connectivities. The correlation profile of this very clustered network is
shown in Fig.\ref{fig_R_60000_triangles}. From  Fig.\ref{fig_R_no_triangles}
one concludes that the correlation  profile in which
connections between hubs are suppressed in favor of connections between
hubs and nodes of low connectivity favors a reduced
number of triangles. If instead nodes with similar connectivities
(including hubs) prefer to connect to each other (the light-colored
area on or around the diagonal in Fig.\ref{fig_R_60000_triangles})
the number of triangles is
typically increased. This in fact can be also demonstrated
analytically. Consider an edge connecting a pair of nodes with connectivities
$K_0$ and $K_1$. The maximal number of triangles containing this edge
is $\min(K_0-1,K_1-1)$. Indeed, in the best case scenario all $K-1$ remaining
neighbors of the smaller connectivity node are also neighbors of the larger
connectivity node. Therefore, given a correlation profile specified by
$N(K_0,K_1)$ - the number of edges connecting nodes with connectivities
$K_0,K_1$ - the absolute maximum number of triangles in the network is given by
$N_{\Delta}^{\max}=\sum_{K_0,K_1}N(K_0,K_1)\min(K_0-1,K_1-1)/6$. Here the factor 1/6
corrects for the fact that in our counting scheme each triangle
would be counted 2 times along each of its three sides. Using
identities $\min(K_0-1,K_1-1)=(K_0-1+K_1-1)/2-|K_0-K_1|$
and $\sum_{K_0,K_1}N(K_0,K_1)(K_0-1)=\sum_{K_0,K_1}N(K_0,K_1)(K_1-1)=
N \langle K(K-1) \rangle$  one finally gets:
\begin{eqnarray}
N_{\Delta}^{\max}&=&{N \langle K(K-1)\rangle \over 6}-\nonumber
\\
&-& {1 \over 6} \sum_{K_0,K_1}N(K_0,K_1) |K_0-K_1| .
\label{eq_max_traingles}
\end{eqnarray}
The first part of this expression corresponds to a hypothetical
situation of the maximal cliquishness in which all neighbors of
every node are connected to each other. It is easy to see
that except for some very special cases of the distribution of connectivities
such maximal cliquishness can never be realized. Indeed, whenever a pair of nodes
of unequal connectivities $K_0,K_1$ are connected to each other the
second term in the Eq. \ref{eq_max_traingles} decreases the
maximal number of triangles. Given the set of node connectivities
$K_i$, one can easily construct the network with the largest possible number of triangles.
One starts by connecting the largest hub node to other nodes in the order of
decreasing connectivities. In the second round of this algorithm one
selects the remaining neighbors of the second largest hub
in the order of decreasing connectivity.
The process continues round by round
until neighbors of all nodes are specified.
When a node reaches its
desired connectivity it will be simply skipped during later rounds of
this algorithm. One can show that the network generated by this
algorithm has the smallest value of $\sum_{K_0,K_1}N(K_0,K_1) |K_0-K_1|$
and the largest number of triangles among all networks with
a given set of node connectivities. In case of the Internet
such network has $63,884$ triangles just below the
$N_{\Delta}^{\max}=64,702$ specified by its correlation profile.
These numbers of triangles are an order of magnitude below
the naive estimate $N \langle K(K-1) \rangle/6 \simeq 690,000$
traditionally used as a normalization factor
in the formula for the clustering coefficient
of a network \cite{watts}. Hence, based on their definition
even the loopiest network with the same node connectivities
as the Internet has a clustering coefficient of only
0.09! For the ``native'' correlation profile of the Internet
Eq. \ref{eq_max_traingles} predicts the maximal number of triangles
close to $24,000$, which sets the observed level of clustering
(6584 triangles) around $27\%$ of its maximal value for this
correlation profile.

In order to check if connectivity correlations visible in the
correlation profile of the internet (Fig.\ref{fig_R_internet})
can fully account for its number of triangles we
generated an ensemble of random networks that preserves not only
connectivities but also the correlation profile
of the complex network.
To this end we used a modification of our main local rewiring
algorithm. There are two principal ways in which this can be done.
In the first scheme,
reminiscent of generating a microcanonical
ensemble in statistical physics,
one allows only for those local rewiring steps that
strictly conserve the number of edges $N(K_0,K_1)$ between
nodes with connectivities $K_0,K_1$.
This is achieved by constraining the selection of pairs of edges
for the rewiring step of Fig.\ref{switch} only to those
connecting nodes
with connectivities $K_0,K_1$, and $K_0,K'_1$.
It is easy to see
that such a local rewiring step strictly conserves $N(K_0,K_1)$.
In practice we softened randomization constraints
by coarse-graining the logarithm of connectivity
to half-decade bins. Using this ``microcanonical algorithm''
we generated an ensemble of
networks with $4132 \pm 75$ loops. The fact that
the number of loops in
the real Internet (6584) is now significantly larger than in
these random networks,
confirms the intuitive notion
that the Internet is indeed characterized by a significant
degree of clustering. We have also found that this
60\%  increase in the level of clustering is equally spread
over the whole spectrum of connectivities.

\begin{figure}[t]
\centerline{\psfig{file=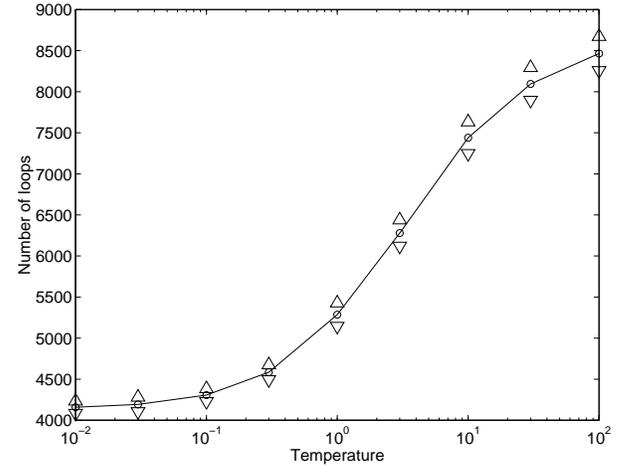,angle=0,width=8cm}}
\vspace{0.5cm}
\caption[]
{The number of loops as a function of temperature observed in
an ensemble of random versions of the Internet generated
by the Metropolis algorithm with the energy function
$H=\sum_{K_0,K_1}[N(K_0,K_1)-N_r(K_0,K_1)]^2/N(K_0,K_1)$.
Upper and lower triangles represent the standard deviation
within an ensemble.}
\label{fig_metropolis_loops}
\end{figure}

As is always the case with microcanonical algorithms one
should worry if the above algorithm is ergodic.
In other words there is no guarantee that
in this algorithm the system does
not get trapped in a disconnected component of the phase space. This
is easily checked by annealing the network
using a {\it canonical} Metropolis algorithm \cite{metropolis} with
an energy function or Hamiltonian, which in our case can be defined as
$H=\sum_{K_0,K_1}[N(K_0,K_1)-N_r(K_0,K_1)]^2/N(K_0,K_1)$,
and sampling networks at a
finite temperature $T$. Local moves lowering the Hamiltonian are
always accepted, while those increasing it by $\Delta H$ are only
accepted with the probability $\exp(-\Delta H/T)$. As seen
in Fig.\ref{fig_metropolis_loops} the above algorithm nicely
extrapolates between the microcanonical algorithm
for small $T$ and the unrestricted local
rewiring algorithm for large $T$.
This confirms that our
microcanonical algorithm is indeed ergodic.

Another conceivable use of the Metropolis algorithm described above
is to generate an artificial
network with a given distribution of connectivities $p(K)$ and
a given correlation profile $R(K_0,K_1)$.
To achieve this one first generates a
seed network with a given $p(K)$, e.g. by the stub
reconnecting algorithm of Ref.\cite{bender,newman}. This network
is first annealed using the Metropolis algorithm with the
energy functional punishing multiple connections between nodes. The resulting
network, containing no multiple connections is
subsequently annealed with another energy functional favoring the
desired correlation profile. This results in an ensemble of
random networks with no multiple connections between nodes
and the desired correlation profile.

In summary we have proposed a general algorithm
to detect characteristic topological features
in a given complex network.
In particular, we introduced the concept of the
{\it correlation profile}, which
allowed us to quantify differences between
different complex networks even when
their connectivity distributions are similar to each other.
Applied to the Internet, this profile identifies
hierarchical features of its structure,
and helps to account for the level of clustering
in this network.

Work at Brookhaven National Laboratory was carried out under Contract
No. DE-AC02-98CH10886, Division of Material Science, U.S.\ Department of
Energy.

\end{document}